\documentclass[runningheads,a4paper]{llncs}

\usepackage{amssymb}
\usepackage{graphicx}
\usepackage{booktabs}
\usepackage[figuresright]{rotating}

\usepackage{url}
\urldef{\mailsa}\path|{nauman.ali, michael.unterkalmsteiner}@bth.se|    
\newcommand{\keywords}[1]{\par\addvspace\baselineskip
\noindent\keywordname\enspace\ignorespaces#1}

\newcommand*\rot{\rotatebox{90}}

\begin{document}

\mainmatter  

\title{Use and evaluation of simulation for software process education: a case 
study}


%
%

\author{Nauman bin Ali
\and Michael Unterkalmsteiner}

%

\institute{Blekinge Institute of Technology\\
\mailsa\\
}

%
%

\toctitle{Lecture Notes in Computer Science}
\tocauthor{Authors' Instructions}
\maketitle

\begin{abstract}

Software Engineering is an applied discipline and its concepts are difficult to
grasp only at a theoretical level alone. In the context of a project management
course, we introduced and evaluated the use of software process simulation (SPS)
based games for improving students' understanding of software development
processes. The effects of the intervention were measured by evaluating the
students' arguments for choosing a particular development process. The arguments
were assessed with the Evidence-Based Reasoning framework, which was extended to
assess the strength of an argument. The results indicate that students generally
have difficulty providing strong arguments for their choice of process models.
Nevertheless, the assessment indicates that the intervention of the SPS game had
a positive impact on the students' arguments. Even though the illustrated
argument assessment approach can be used to provide formative feedback to
students, its use is rather costly and cannot be considered a replacement for
traditional assessments.
\keywords{Software Engineering education, Software process simulation, project 
management, argument evaluation}
\end{abstract}

\section{Introduction}
The Software Engineering (SE) discipline spans from technical aspects, such as
developing techniques for automated software testing, over defining new
processes for software development improvement, to people-related and
organizational aspects, such as team management and leadership. 
This is evident in the software development process, which is  \textit{``the
coherent set of policies, organizational structures, technologies, procedures,
and artifacts that are needed to conceive, develop, deploy, and maintain a
software product''} \cite{Fuggetta2000}. This breadth of topics encompassed here
makes education in SE challenging as the interaction of the different
disciplines cannot be exclusively taught on a theoretical level, but must also
be experienced in practice. As such, SE education needs to identify means to
prepare students better for their tasks in
industry~\cite{lethbridge_improving_2007}.

However, the complexity and dynamism of software processes makes it difficult to
illustrate the implications of the chosen development process on the outcomes of
a project. Students will have to undertake multiple iterations of developing the
same project using different software development processes to understand the
various processes and their implication on the project attributes
\cite{Wangenheim2009}. Such repetitions are however impractical because of the
time and cost involved. To overcome this shortcoming software process simulation
(SPS) has been proposed as a means of SE education. SPS is the numerical
evaluation of a computerized-mathematical model that imitates the real-world
software development process behavior \cite{Kellner1999}. It has been found to
be useful in SE education as a complement to other teaching methods e.g. in
combination with lectures, lab sessions and projects
\cite{Navarro2007,Wangenheim2009}.

In this paper we motivate, illustrate and evaluate how a targeted change was
introduced in the graduate-level \textit{Applied Software Project Management
(ASPM)} course. The course aims to convey to students in a hands-on manner how
to prepare, execute and finalize a software project. In previous instances of
the course, we have observed that students encounter difficulties in choosing an
appropriate software development process and in motivating their choice. We
hypothesize that the students lack experience of different software development
processes, and lack therefore the analytical insight required to choose a
process appropriate for the characteristics of the course project. We study our
hypothesis by exposing students to software process simulations (SPS) and by
evaluating thereafter the argumentative strength for choosing/discarding a
particular process.

There are three major contributions in this paper. First, a review of frameworks
for evaluating argumentative reasoning was updated to cover more recent
research. Secondly the framework relevant for evaluating arguments in the
context of SE was selected and adapted. Thirdly, independent of the creators of 
SimSE, we used it in the context of
an active course instead of a purely experimental setting, and evaluated its 
effect indirectly, in terms of students' understanding of software development 
processes.

The remainder of the paper is structured as follows: Section
\ref{sec:background} summarizes the relevant work on the topic of SPS in SE
education. Section \ref{sec:researchdesign} presents the context of the study,
research questions, data collection and analysis methods. Section
\ref{sec:results} presents the results, Section~\ref{sec:discussion} revisits 
the research questions based on the findings and Section \ref{sec:conclusion} 
concludes the paper.

\section{Background and Related Work}
\label{sec:background}
In this section, we briefly discuss the two overarching themes in this study: 
SPS based education and evaluation on scientific argumentation. 
\subsection{SPS in SE education}
SPS provides an alternative to manipulation of the actual software process by
providing a test-bed for experimentation with realistic considerations. Compared
to static and analytical models, SPS achieves this because of its ability to
capture the underlying complexity in software development by representing
uncertainty, dynamic behavior and feedback/feed-forward mechanisms
\cite{Kellner1999}.

Since the initial proposal of SPS its potential as a means of
education and training was recognized \cite{Kellner1999}. Some of the claimed
benefits of SPS for SE education include: increased interest in SE project
management \cite{Pfahl2003}, motivation of students \cite{Drappa2000}, and
effective learning \cite{Rodriguez2006}. It can facilitate understanding by
experiencing different processes with certain roles (e.g. as a as a software
manager making decisions in software development, which would not have been
possible in an academic context without SPS \cite{Zapalska2012}).

Navarro and Hoek \cite{Navarro2007} evaluated the experience of students playing
SPS based games for SE education. They found that the SPS based teaching is
applicable for various types of learners as it aligns well with objectives of a
multitude of learning theories. For example, it encourages exploratory learning
by experimenting, emphasizes learning by doing and through failure, and by
embedding in a context that resembles the real-world use of the phenomenon of
interest.

Wangenheim and Shull  \cite{Wangenheim2009}, in a systematic literature review
of studies using SPS for SE education, found that the two most frequent aims in
such studies are \textit{``SE Project Management''} and \textit{``SE process''}
knowledge \cite{Wangenheim2009}. 
They also found that in most of the existing research, subjective feedback was
collected after the students had used the game \cite{Wangenheim2009}. Similarly,
they reported that it was difficult to evaluate the effectiveness of SPS
interventions because a majority of the articles do not report the
\textit{``expected learning outcome and the environment in which students used
the game''} \cite{Wangenheim2009}.

These findings motivated our choice to have a simulation based intervention in
the course as the two major learning objectives for the course are related to
project and process management. The context is described in Section
\ref{sec:context}. Furthermore, adhering to the recommendation that is based on
empirical studies \cite{Wangenheim2009}, we used SPS to target a
\textit{``specific learning need''} of the students, i.e. to improve the
understanding and implications of a software development lifecycle process.
SimSE was the chosen platform due to a stable release, good graphical
user-interface and good feedback from earlier evaluations \cite{Navarro2007}.
Unlike the existing evaluations of SimSE, in this study, we took an indirect
approach to see if the simulation based intervention had the desired impact. We
looked at the quality of arguments for the choice of the lifecycle process in
the student reports without explicitly asking them to reflect on the SPS game.




\subsection{Evaluating scientific argumentation}\label{sec:evalargs}
Argumentation is a fundamental driver of the scientific discourse, through which
theories are constructed, justified, challenged and
refuted~\cite{erduran_tapping_2004}. However, scientific argumentation has also
cognitive values in education, as the process of externalizing one's thinking
fosters the development of knowledge~\cite{erduran_tapping_2004}. As students
mature and develop competence in a subject, they pass through the levels of
understanding described in the SOLO taxonomy~\cite{biggs_evaluating_1982}. In
the taxonomy's hierarchy, the quantitative phase (unistructural and
multistructural levels) is where students increase their knowledge, whereas in
the qualitative phase (relational and extended abstract levels) students deepen
their knowledge~\cite{biggs_teaching_2007}. The quality of scientific
argumentation, which comprises skills residing in higher levels of the SOLO
taxonomy, is therefore a reflection of the degree of understanding and
competence in a subject.

As argumentation capability and subject competence are intrinsically related, 
it is important to find means by which scientific argumentation in the 
context of education can be evaluated. Sampson and 
Clark~\cite{sampson_assessment_2008} provide a review of frameworks developed 
for the purpose of assessing the nature and quality of arguments. They analyze 
the studied frameworks along three dimensions of an 
argument~\cite{sampson_assessment_2008}:
\begin{enumerate} 
 \item Structure (i.e., the components of an argument)
 \item Content (i.e., the accuracy/adequacy of an arguments components when 
evaluated from a scientific perspective)
 \item Justification (i.e., how ideas/claims are supported/validated within an 
argument)
\end{enumerate}

We used the same criteria to update their review with newer frameworks for
argument evaluation. This analysis was used to select the framework appropriate 
for use in this study.


\section{Research design}
\label{sec:researchdesign}
%

\subsection{Context}\label{sec:context}
The objective of the Applied Software Project Management (ASPM) course is to
provide students with an opportunity to apply and sharpen their project
management skills in a sheltered but still realistic environment.
Students participating in ASPM typically\footnote{ASPM is also an optional
course in the curriculum for students from computer science and civil
engineering programs} have completed a theory-building course on software
project management, which includes an introduction to product management,
practical project management guided by the Project Management Body of
Knowledge~\cite{_guide_2004}, and an excursion to leadership in project
teams~\cite{hersey_management_2001}.

\begin{figure*}[htbp]
\centering
\includegraphics[scale=0.8]{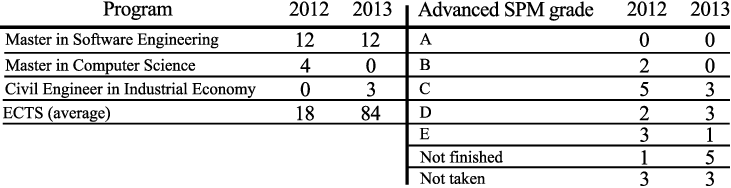}
\caption{Student demographics from 2012 (without intervention) and 2013 (with 
SPS intervention) of the Applied SPM course}
\label{fig:studentdemographics} 
\end{figure*}

Figure~\ref{fig:studentdemographics} shows the student characteristics of the
two course instances that were studied. In 2012, without SPS intervention, 16
students participated in total, having accumulated on average 18 ECTS points at
the start of the course. In 2013, with the SPS intervention, 15 students
participated in total, having accumulated on average 84 ECTS points at the start
of the course. In both course instances, three students did not take the theory
course on software project management (Advanced SPM). The major difference
between the two student groups is that in 2013, considerably more students did
not successfully complete the Advanced SPM course. The higher ECTS average in
2013 can be explained by the participation of three Civil Engineering students
who chose Applied SPM at the end of their study career while SE and
Computer Science students chose the course early in their studies.

\begin{figure*}[htbp]
\centering
  \includegraphics[scale=0.5]{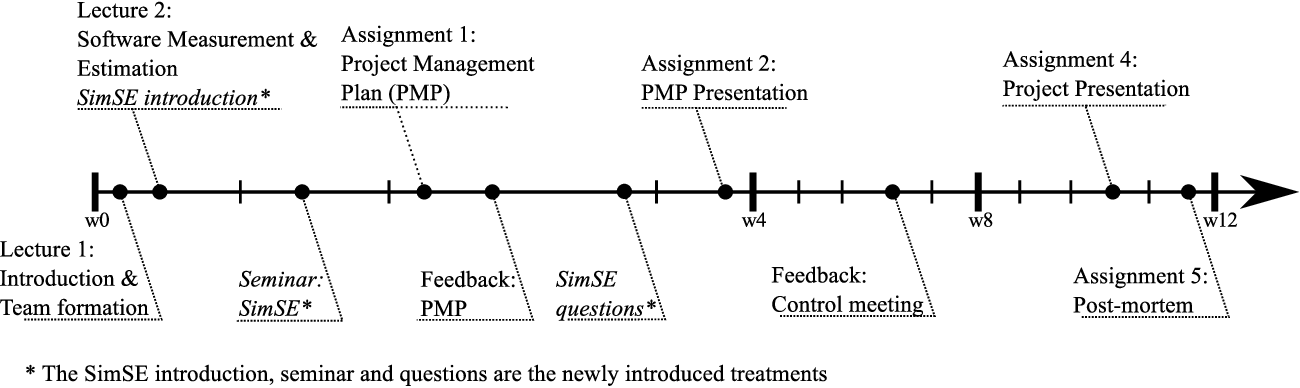}

\caption{ASPM course time-line with events}
\label{fig:coursetimeline} 
\end{figure*}

The course follows the three months schedule shown in 
Figure~\ref{fig:coursetimeline}, which illustrates also the planned 
interactions between students and instructors. The introduced modifications are 
shown in italics and further discussed in Section~\ref{sec:instr}. Students 
are expected to work 200 hours for this course, corresponding to a 20 
hours/week commitment.


The course has five assignments but Assignment 1 and 5 are important for this
study (see Figure~\ref{fig:coursetimeline}). Assignment 1 consists of delivering
a project management plan (PMP) where students also report the choice and
rationale for a software process they will use.  The teams receive oral feedback
and clarifications on the PMP during the same week. The course concludes with a
presentation where project teams demo their developed products. In Assignment 5,
the students are asked to individually answer a set of questions that, among
other aspects, inquiry their experience with the used software process in the
project.

\subsection{Research questions}\label{sec:researchquestions}
The posed research questions in this study are:
\begin{description}
  \item RQ1: How can improvement in software development process understanding 
be assessed?
  \item RQ2: To what extent can process simulation improve students' 
understanding of software development processes?

\end{description}

With RQ2, we investigate whether process simulation has a positive impact on
students' understanding of development processes. Even though studies with a
similar aim have already been conducted (c.f. \cite{Pfahl2003}), experiments in 
general are prone to the Hawthorne effect~\cite{campbell_hawthorne_1995}, where 
subjects under study modify their behavior knowing that they are being 
observed. Similar limitations can be observed in earlier evaluations of SimSE 
where \textit{``the students were given the assignment to play three SimSE 
models and answer a set of questions  concerning the concepts the models are 
designed to teach''} \cite{Navarro2007}. 
Hence we introduce process
simulation as an additional teaching and learning activity into a course whose
main purpose is \emph{not} to teach software development processes. Furthermore,
we do not modify the requirements for the graded deliverables. Formally, we 
stated the following hypotheses:
\begin{itemize}  
 \item[$H_0$:] There is no difference in the students' understanding of process 
models in course instances 2012 and 2013.
 \item[$H_a$:] There is a difference in the students' understanding of process 
models in course instances 2012 and 2013.
\end{itemize}

Due to the subtle modifications in the course, we needed to identify 
new means to evaluate the intervention, measuring the impact of introducing 
process simulation on students' understanding of development processes. In order 
to answer RQ1, we update the review by Sampson and 
Clark~\cite{sampson_assessment_2008} with two more recent frameworks proposed by 
Reznitskaya et al.~\cite{reznitskaya_measuring_2009} and Brown et 
al.~\cite{brown_evidence-based_2010}, select the framework that provides the 
strongest argument evaluation capabilities, and adapt it to the
software engineering context.

In order to answer RQ2, we apply the chosen argument evaluation framework on
artifacts delivered by project teams and individual students which did receive
the treatments shown in Figure~\ref{fig:coursetimeline} and on artifacts
delivered in the previous year.

\subsection{Instrumentation and data collection}\label{sec:instr}
\begin{figure}[htbp]
\centering
  \includegraphics[scale=0.5]{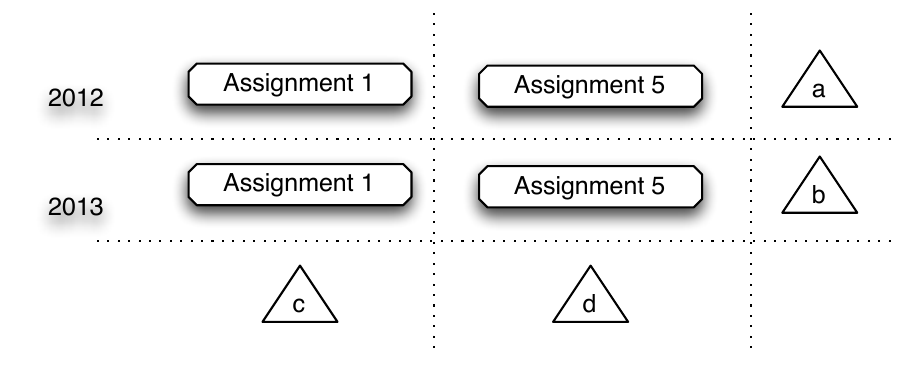}

\caption{Design to evaluate the impact of the SPS intervention}
\label{fig:studyDesign} 
\end{figure}

\textit{Assignment 5: Post mortem}, as shown in Figure~\ref{fig:coursetimeline},
is an individual assignment where students had to motivate and reflect on their
choice of software process model selected in their projects. This assignment is
used to evaluate the influence of SimSE on the student's understanding of the
software processes. A baseline for typical process understanding of students
from the course was established by evaluating \textit{Assignment 5} from year
2012 and it was compared to the evaluation results of \textit{Assignment 5} from
year 2013. To supplement the analysis we also used \textit{Assignment 1: Project
Management Plan (PMP)} (which is a group assignment) from both years. The design
for the study is shown in Figure~\ref{fig:studyDesign}. Where deltas
\textit{`a'} and \textit{`b'} are changes in understanding between the
Assignments 1 and 5 within a year. While deltas \textit{`c'} and \textit{`d'}
represent changes across the years for Assignment 1 and 5 respectively.

For the evaluation of assignments, we used the EBR
framework~\cite{brown_evidence-based_2010}. Other frameworks considered and the
reasons for this choice are summarised in Section~\ref{sec:frameworks}. Once the
framework had been adapted, first it was applied on one assignment using
\textit{``Think-aloud protocol''} where the authors expressed their thought
process while applying the evaluation instrument. This helped to identify
ambiguities in the instrument and also helped to develop a shared understanding 
of
it. A pilot of the instrument was done on two assignments where the authors
applied it separately and then compared and discussed the results. Both authors
individually coded all the assignments and then the codes were consolidated with
consensus. The results of this process are presented in
Section~\ref{sec:application}.

\vspace{-0.4cm}
\vspace{-0.99cm}

\subsection{Limitations}\label{sec:limitations}
The assignments were retrieved from the Learning Management System and personal 
identification of students was replaced with a unique identifier to hide their 
identity from the authors. This was done to avoid any personal bias that the 
authors may have towards the students as their teachers, in this and other 
courses. Furthermore, to ensure an unbiased assessment both the overall grades 
of students and their grades in the assignments were hidden from the authors 
when the assessment instrument was applied in this study.

To avoid any bias introduced by asking questions directly about the 
intervention of process simulation, and to have a relative baseline for 
assignments from 2012, we did not change the assignment descriptors for the 
year 2013. Thus we tried to measure the effect of the intervention indirectly 
by observing the quality of argumentation without explicitly asking students to 
reflect on the experience from simulation based games.

The intervention was applied in a post-graduate course, rendering 
experiment-like control of settings and variables impossible. Among other 
factors, 
any difference in results could purely be because of the different set of 
students taking the course in the years 2012 and 2013. However, as discussed in 
Section \ref{sec:context} the groups of students were fairly similar thus the 
results are comparable.  Small number of students is also a limitation of this 
study especially to draw any statistical inference with high confidence. 

Similarly, by having the students fill out questions about the various 
simulation based games we tried to ensure that students have indeed played the 
games. However, we have no way of ensuring that the students did indeed play 
the games individually and not share the answers with each other. This 
limitation could weaken the observable effect of the intervention.




\section{Results}
\label{sec:results}
In this section we report the two main results of our study. In 
Section~\ref{sec:reviewupdate} we review two argument evaluation approaches and 
classify them according to the framework presented in 
Section~\ref{sec:evalargs}. Then we choose one argument evaluation approach and 
adapt it to our goals (Section~\ref{sec:frameworks}), and apply it to students 
arguments on choosing a particular process model for their project 
(Section~\ref{sec:application}). The data (argument coding and quantification) 
is available in the supplementary material to this paper, available at 
http://www.bth.se/com/mun.nsf/pages/simeval.

\subsection{Review update}\label{sec:reviewupdate}

In Table~\ref{tab:sw} we summarize Sampson and 
Clark's~\cite{sampson_assessment_2008} analysis w.r.t. the 
support various frameworks provide to assess structure, content and 
justification of an argument. In the rest of the section we report the 
classification of two newer frameworks as an extension to their review.

\begin{table*}[htbp]
\caption{Strengths and weaknesses of argument assessment frameworks}
\label{tab:sw}
\footnotesize
\centering
\scalebox{0.9}{
\begin{tabular}{lccc}
\toprule
Framework & Structure & Content & Justification \\
\midrule
\midrule
\multicolumn{4}{l}{Domain-general} \\
Toulmin~\cite{toulmin_uses_1958} & strong & weak & weak \\
Schwarz et al.~\cite{schwarz_construction_2003} & strong & moderate & moderate 
\\
\midrule
\multicolumn{4}{l}{Domain-specific} \\
Zohar and Nemet~\cite{zohar_fostering_2002} & weak & moderate & strong \\
Kelly and Takao~\cite{kelly_epistemic_2002} & strong & weak & strong \\
Lawson~\cite{lawson_nature_2003} & strong & weak & strong \\
Sandoval~\cite{sandoval_conceptual_2003} & weak & strong & strong \\
\bottomrule
\end{tabular}
}
\end{table*}

Brown et al.~\cite{brown_evidence-based_2010} propose with the Evidence-based 
Reasoning (EBR) framework an approach to evaluate scientific reasoning that 
combines Toulmin's argumentation pattern~\cite{toulmin_uses_1958} with Duschl's 
framework of scientific inquiry~\cite{duschl2003assessment}. This 
combination proposes scientific reasoning as a two-step process in which a 
scientific approach to gather and interpret data results in rules 
that are applied within a general framework of 
argumentation~\cite{brown_evidence-based_2010}. 

\begin{figure*}[htbp]
\centering
  \includegraphics[scale=0.5]{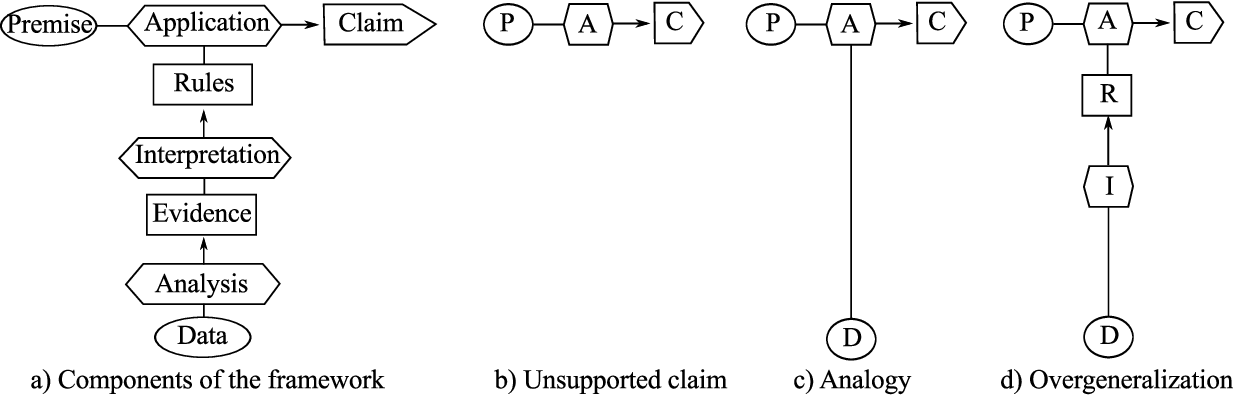}
\caption{The Evidence-Based Reasoning framework (a) and different degrees of 
argument sophistication (b-d) (adapted from Brown et 
al.~\cite{brown_evidence-based_2010})}
\label{fig:ebr} 
\end{figure*}

Figure~\ref{fig:ebr}a shows the structure of the framework, consisting of 
components that should be present in a strong scientific argument. The strength 
of an argument can be characterized by the components present in the argument. 
For example, in an unsupported claim (Figure~\ref{fig:ebr}b), there are no 
rules, evidences or data that objectively support the claim. An analogy 
(Figure~\ref{fig:ebr}c) limits the argumentation on supporting a claim 
only with instances of data, without analyzing and interpreting the data. In 
an overgeneralization (Figure~\ref{fig:ebr}d), analysis and formation of a 
body of evidence is omitted and data is interpreted directly to formulate rules 
(theories, relationships) that are not supported by evidence.

The EBR was not designed for a specific scientific 
context~\cite{brown_evidence-based_2010} and we classify it therefore as a 
domain-general framework. It provides strong support for evaluating arguments 
along structure (i.e. the components of an argument) and justification (i.e. 
how claims are supported within an argument) dimensions. However, solely 
identifying rules and evidences components in an argument does not provide an 
assessment of the arguments content, i.e. the adequacy of argument components. 
As such, the framework provides the tools to identify components of content 
(rules and evidences), but no direct means to assess the content's quality. 
Therefore, we rate the framework's support for the content dimension as 
moderate.

Reznitskaya et al.~\cite{reznitskaya_measuring_2009} propose a domain-specific
framework for evaluation of written argumentation. They proposed and compared 
two methods to implement this framework: 
\begin{enumerate} 
	\item Analytical method, which is a data driven approach where individual 
statements are coded and their relevance to the main topic is judged, 
categories are derived from these codes (deciding about these categories will 
be based on the theoretical and practical significance in the domain). Next 
the report is evaluated on five sub-scales which cover aspects from: 
number of arguments made, types of previously identified categories of 
arguments covered in the report, opposing perspectives considered, number of 
irrelevant arguments and the use of formal aspects of discourse. 

	\item Holistic method takes a rubric based approach attempting to provide a 
macro level assessment of the arguments.
\end{enumerate}
In essence, the framework has no explicit focus on the components of an 
argument and only indirectly covers the aspects of structure while creating the 
instrument. The fundamental building block of the evaluation framework is the 
analytical coding process where both the content and justification are 
considered. Content (accuracy and adequacy) is only assessed by 
identifying the relevance of the argument to the topic. Justification is 
covered indirectly in the variety of argument categories identified in the 
reports. However, all argument categories are given equal weight in scoring. 
Therefore, we rate the framework support for the structure as weak, and for 
content and justification as moderate.
\subsection{Selection and adaptation of an argument evaluation framework} 
\label{sec:frameworks}



Based on the analysis of the reviewed 
frameworks, summarized in Table~\ref{tab:sw}, and the updated review presented 
in Section~\ref{sec:reviewupdate}, we decided to use 
the EBR framework~\cite{brown_evidence-based_2010}. We chose a domain-general 
over a domain-specific framework due to our preference of customizing generic 
principles to a specific context. The alternative, to construct an assessment 
instrument completely inductively from the particular domain and data, as for 
example in Reznitskaya et al.~\cite{reznitskaya_measuring_2009}, would imply a 
relative assessment approach, weakening the evaluation of the intervention. 
Furthermore, a domain-generic approach allows us to re-use the assessment 
instrument with minor adaptations, lowering the application cost by keeping the 
general approach intact. 

Table~\ref{tab:ebrapplication} shows an example argument with the EBR 
components one can identify in written statements. This example illustrates 
what we would expect in a strong argument: a general rule on a process model 
is applied on a premise that refers to the specific circumstances of the 
students' project, justifying their claim that the selected model was the best 
choice. The rule is supported by evidence (a reference to a scientific study) 
and by an experience from the project that creates a relationship between short 
development cycles and late changes. The evidence is supported by data from the 
project.

\begin{table*}[htbp]
\caption{Example application of the EBR on an ideal argument}
\label{tab:ebrapplication}
\footnotesize
\centering
\scalebox{0.9}
{
\begin{tabular}{lp{10cm}}
\toprule
Component & Statement \\
\midrule
\midrule
Premise & The requirements for our product are not that clear and likely to 
change. \\
Claim & eXtreme Programming (XP) was the best choice for our project. \\
Rule & XP embraces continuous change by having multiple short development 
cycles. \\
Evidence & Reference to Beck~\cite{beck1999}; customer changed user interaction 
requirements six weeks before the delivery deadline but we still delivered a 
working base product; \\
Data & Seven change requests to initially stated requirements; four 
requirements were dropped since customer was satisfied already; \\
\bottomrule
\end{tabular}
}
\end{table*}

The EBR framework enables a fine-grained deconstruction of arguments into 
components. The price for this strength on the structural and justification 
dimension is a moderate support for assessing content (see 
Section~\ref{sec:reviewupdate}). Even though the overall argument content can 
be judged to some extent by the interplay between the argument components, 
domain-specific knowledge is nevertheless required to judge whether individual 
statements are accurate. Looking at Table~\ref{tab:ebrapplication}, the rule 
component in particular requires knowledge on XP in order to decide whether the 
statement is accurate or not. Hence we assess the argument content by qualifying 
a rule as sound/unsound, given the stated premise, claim, evidence and data, 
based on our domain knowledge on process models. Concretely, we declare an 
argument for a claim as:
\begin{itemize} 
  \item Sound
  \begin{itemize} 
    \item If the stated rule is backed by evidence/data \emph{and} is pertinent 
for the premise (strong).
    \item In arguments with no corroborative evidence (weak): If the stated 
rule is in compliance with literature and/or the assessors understanding of the 
topic \emph{and} is pertinent for the premise.
  \end{itemize}
  \item Unsound
  \begin{itemize} 
    \item If an argument does not fulfill either of the above two criteria.
  \end{itemize}
\end{itemize}
 
\subsection{Application of the chosen framework}\label{sec:application}

In this section we illustrate the results of applying the EBR framework on the 
students' arguments for choosing/rejecting a particular process model for their 
project. We coded statements according to the EBR frameworks' components of an 
argument: a premise (P), rule (R), evidence (E), data (D). We also noted when 
components are used atomically or are combined into pairs or triples of 
components to form a coherent argument. Based on this codification, we 
evaluated the overall content of the argument (unsound / weak / strong) by 
following the rules established in Section~\ref{sec:frameworks}. The claim of 
the argument, in principle constant and only changing in sign, was that a 
particular process model is / is not the best choice for the project. 

Table~\ref{tab:ebrpmp} shows the results in terms of the argument 
component frequencies encountered in the students' project plans from 2012 
(without intervention) and 2013 (with SPS intervention). For example, in Plan 
\#1 we identified 8 premises (P), 4 premise-rule (PR) pairs and 1 
premise-evidence (PE) pair. We expected to find some premise-rule-evidence 
(PRE) triples as they would indicate that students can motivate their choice 
by examples, e.g. by referring to scientific literature, to experience from 
previous projects or from playing SimSE. However, the results clearly indicate 
a tendency for students to create overgeneralizing arguments (premise-rule 
pairs). Looking at the argument content, we identified no strong arguments 
(lack of evidence component) and, in proportion to weak arguments, a rather 
large number of unsound arguments, indicating a lack of understanding of 
process models and their properties.

\begin{table*} 
\caption{Frequencies of identified argument components and argument content 
strength in project plans for choosing a particular process model}
\label{tab:ebrpmp}
\footnotesize
\centering
\scalebox{0.9}
{
\begin{tabular}{p{1cm}cccccccccc}
\toprule
Year & Plan\# & P & PR & PE & PRE & RE & R & Unsound & Weak & Strong \\
\midrule
\midrule
& 1& 8 & 4 & 1 & 0 & 0 & 0 & 2 & 2 & 0 \\
2013 & 2 & 9 & 7 & 0 & 0 & 0 & 3 & 4 & 6 & 0 \\
& 3 & 3 & 1 & 0 & 0 & 1 & 0 & 1 & 1 & 0 \\
& \textbf{Sum} & \textbf{20} & \textbf{12} & \textbf{1} & \textbf{0} & 
\textbf{1} & \textbf{3} & \textbf{7} & \textbf{9} & \textbf{0} \\
\midrule
& 4 & 10 & 3 & 0 & 0 & 0 & 1 & 1 & 3 & 0 \\
& 5 & 5 & 7 & 0 & 0 & 1 & 1 & 3 & 6 & 0 \\
2012 & 6 & 6 & 3 & 0 & 0 & 0 & 0 & 0 & 4 & 0 \\
& 7 & 2 & 1 & 0 & 0 & 0 & 2 & 1 & 1 & 0 \\
& \textbf{Sum} & \textbf{23} & \textbf{14} & \textbf{0} & \textbf{0} & 
\textbf{1} & \textbf{4} & \textbf{5} & \textbf{14} & \textbf{0} \\
\bottomrule
\end{tabular}
}
\end{table*}

After assessing the project plans, which were a group assignment, we applied 
the EBR framework on the project post-mortems that were handed in individually 
by the students (13 in 2013 and 12 in 2012). Table~\ref{tab:ebrpm} illustrates 
the results, showing the frequencies of identified argument components and 
component combinations. Observe that, in contrast to the post-mortem, we 
identified more combinations of components, e.g. premise-data (PD) or 
premise-evidence-data (PED) triples. For both years it is evident that students 
reported more justification components (evidence and data) in the post-mortem 
than in the project plan. This is expected as we explicitly asked to provide 
supporting experience from the conducted project. 

%

\vspace{-0.4cm}
\begin{table} 
\caption{Frequencies of identified argument components and argument content 
strength in project post-mortems for choosing a particular process model.}
\label{tab:ebrpm}
\footnotesize
\centering
\begin{tabular}{p{1cm}p{0.4cm}p{0.4cm}p{0.4cm}p{0.4cm}p{0.4cm}p{0.4cm}p{0.4cm}p{
0.4cm}p{0.4cm}p{0.4cm}p{0.4cm}p{0.4cm}p{0.4cm}p{0.4cm}p{0.4cm}p{0.4cm}p{0.4cm}p{
0.4cm}}
\toprule
Year & \rot{Premise} & \rot{Premise-Rule} & \rot{Premise-Evidence} & 
\rot{Premise-Rule-Evidence} & \rot{Rule-Evidence} & \rot{Rule} & \rot{Evidence} 
& \rot{Data} & \rot{Evidence-Data} & \rot{Premise-Data} & \rot{Rule-Data} & 
\rot{Rule-Evidence-Data} & \rot{Premise-Rule-Data} & \rot{Premise-Evidence-Data} 
& \rot{Premise-Rule-Evidence-Data} & \rot{\textbf{Unsound Argument}} & 
\rot{\textbf{Weak Argument}} & \rot{\textbf{Strong Argument}} \\
\midrule
\midrule
2012 & 5 & 2 & 3 & 4 & 14 & 11 & 16 & 4 & 8 & 4 & 2 & 3 & 0 & 1 & 2 & 17 
& 14 & 7 \\
2013 & 30 & 5 & 5 & 9 & 17 & 20 & 9 & 16 & 9 & 3 & 9 & 6 & 2 & 1 & 0 & 12 
& 42 & 14 \\
\bottomrule
\end{tabular}
\end{table}
\vspace{-0.4cm}

\section{Analysis and revisiting research questions}\label{sec:discussion}

\subsection{Assessing software development process understanding (RQ1)}
With support from the EBR framework we decomposed students' arguments to a 
degree that allowed us to pinpoint the weaknesses of their development process 
model understanding. Looking at the frequencies in Table~\ref{tab:ebrpm}, we 
can observe that:
\vspace{-0.4cm}
\begin{itemize}  \itemsep -1pt 
 \item For both years, a relatively large number of standalone argument
components were identified (e.g. single premise, rule and data components in 
2013 and evidence components in 2012). A standalone argument component 
indicates a lack of a coherent discussion, a concatenation of information 
pieces that does not create a valid argument for a specific claim. There are 
exceptions, e.g. PM\#8 and PM\#24 (see supplementary material), which is also 
expressed in a strong argument content rating.

 \item Looking at the argument component combinations that indicate a strong 
argument (i.e. that contain a premise, a rule, and evidence or data), we can 
observe that assignments containing weak arguments outnumber assignments 
with strong arguments in both years.
\end{itemize}
\vspace{-0.1cm}

The detailed analysis of arguments with the EBR framework could also help to 
create formative feedback to students. For example, in PM\#5, the student 
reported 7 data points on the use of SCRUM but failed to analyze this data, 
creating evidence (describing relationships of observations) and 
connecting them to rules. On the other hand, we identified several assignments 
with the premise that the requirements in the project are unknown or change 
constantly (using this premise to motivate the selection of an Agile process). 
However, none of the assignment reports data on change frequency or volatility 
of requirements, weakening therefore the argument for choosing an Agile process.

Given these detailed observations one can make by applying the EBR framework we 
think it is a valuable tool for both assessing arguments and to provide 
feedback to students. However, this power comes at a price. We recorded 
coding times between 10 and 30 minutes per question, excluding any 
feedback formulation that the student could use for improvement. Even if coding 
and feedback formulation efficiency could be increased by tools and routine, 
one has to consider the larger effort this type of assessment requires 
compared to other means, e.g. rubrics~\cite{barney_improving_2012}.

%

\subsection{Impact of SPS on students' understanding of software development 
processes (RQ2)}

The only difference in how the course was conducted in 2013 compared to 2012 
was the use of SimSE simulation based software process games. Besides 
the limitation of this study (as discussed in Section~\ref{sec:limitations}) 
improvements in the students' understanding can be seen as indications of 
usefulness of SimSE based games for software process education. 

In order to evaluate students' understanding, we measured the quality of 
argumentation for the choice of the software process. Concretely, we evaluated 
the content of the student reports by using the strength (classified as strong, 
weak and unsound) of an argument as an indicator. To test the hypotheses 
stated in Section~\ref{sec:researchquestions}, we used the chi-square test 
of independence~\cite{sheskin_handbook_2000} and rejected $H_0$ at a confidence 
level of $\alpha < 0.05$.

For the project management plan (Table~\ref{tab:ebrpmp}), which was a group 
assignment and was delivered at the beginning of the course, the observed 
frequency of strong, weak and unsound arguments did not differ 
significantly between 2012 and 2013. Hence we cannot reject $H_0$ for the 
project management plan.

For the project post-mortem (Table~\ref{tab:ebrpm}), which was an individual 
assignment and was delivered at the end of the course, the observed frequency 
of strong, weak and unsound arguments did differ significantly between 2012 and 
2013 ($chi-squared=8.608$, $df=2$, $p-value=0.009$). Hence we can reject $H_0$ 
for the project post-mortem and accept that there is a difference in process 
model understanding of students in course instances 2012 and 2013. However, 
since this difference only materialized at the end of the course, after the 
project has been conducted, the improved understanding cannot be attributed to 
the software process simulation game alone as discussed in the limitations of 
this study (Section~\ref{sec:limitations}).


Another indirect observation that shows a better understanding of the process
model is reflected in the choice of the process model in the context of the
course project (with small collocated teams, short fixed duration for project
etc.). Compared to 2012 where most of the groups took plan driven, document
intensive process models (two groups chose an incremental development model, one
group chose Waterfall and only one chose Scrum), in the year 2013 all groups
chose a light-weight, people centric process model that is more pertinent to the
given context.

\section{Conclusion}
\label{sec:conclusion}
The EBR framework enabled decomposition of arguments into distinct
parts, which ensured an objective evaluation of the strength of the arguments
in student reports. This assessment allowed us to gauge students'
understanding of software development processes.

The indications reported in this study (from use of software process simulation
in an active course) adds to the confidence in evidence reported in earlier
empirical studies in controlled settings. Given the potential gains as seen in
this study, and relative maturity, user interface and decent documentation of
SimSE, the minor additional cost of including it in a course to reinforce
concepts already learned was well justified.

As future work, we intend to do a longitudinal study where more data is 
collected over the next few instances of the course. 

\vspace{-0.3cm}
\bibliographystyle{abbrv}
\bibliography{main}

\end{document}